\documentclass[nofootinbib,aps,prl,groupedaddress]{revtex4}
\usepackage{amsfonts,amssymb}
\usepackage{amsthm}
\usepackage{tikz}
\usetikzlibrary{calc}
\usetikzlibrary{arrows}
\usepackage{tikz-3dplot}
\usepackage{color} 
\usepackage[fleqn]{amsmath}
\usepackage{amsthm}
\usepackage{amssymb}
\usepackage{amsfonts}
\usepackage{bbm}
\usepackage{epsfig}
\usepackage{times}
\usepackage{hyperref}
\usepackage{bm}
\usepackage{float} 
\usepackage{appendix} 
\usepackage{mathrsfs}
\usepackage{appendix}

\usepackage{graphicx}

\DeclareGraphicsExtensions{.png}

\newcommand{\comment}[1]{}

\theoremstyle{plain}

\theoremstyle{definition}

\begin{document}

\title{Replication Ethics} 

\author{Adrian \surname{Kent}} \affiliation{Centre for
	Quantum Information and Foundations, DAMTP, Centre for Mathematical
	Sciences, University of Cambridge, Wilberforce Road, Cambridge, CB3
	0WA, U.K.}  \affiliation{Perimeter Institute for Theoretical
	Physics, 31 Caroline Street North, Waterloo, ON N2L 2Y5, Canada.}
	\date{\today}
	
\newpage
	
\begin{abstract} 
Suppose some future technology enables the same consciously experienced human life to 
be repeated, identically or nearly so, $N$ times, in
series or in parallel.  Is this roughly $N$ times as valuable as enabling the
same life once, because each life has value and values are additive?   
Or is it of roughly equal value as enabling the life once, because
only one life is enabled, albeit in a physically unusual way? 
Does it matter whether the lives are contemporaneous or successive? 
We argue that these questions highlight a 
hitherto neglected facet of population ethics that may
become relevant in the not necessarily far distant future. 
\end{abstract}
\maketitle
	
\section{Introduction}

\subsection{Future population ethics}

Relatively few people are strict utilitarians.
Indeed, most probably don't believe there is any generally applicable
way of deciding ethical questions based on a utility function
calculus.  Nonetheless, many think that some form of 
utilitarian reasoning is sometimes valid, or at least
that it gives a simple and useful policy guide for
some practical issues. 
Killing two random unknown people as a consequence of some policy choice 
seems worse than killing one, ceteris paribus. 
For most people, it seems roughly twice as bad.   

More complex examples such as trolley problems \cite{sep-doing-allowing}, in which strict utilitarianism 
suggests that costs to one group may be offset against benefits
to another, cause much more controversy. 
Utilitarian reasoning is more controversial still 
when it comes to questions of existential risk and the future
of life.    
For example, Parfit's Repugnant Conclusion \cite{parfit1984reasons} that
$M$ lives barely worth living are preferable to $M'$ fulfilled
and happy lives, for sufficiently large $M \gg M'$, highlights
a clash between utilitarianism and many people's intuitions,
including Parfit's.    
Nozick's Utility
Monster \cite{nozick1974anarchy}, which gains such great utility from
resources that utilitarianism suggests everyone else's interests
should be sacrificed in its maw, similarly troubles many.   
To sustain the intuitions that the conclusion is incorrect as well as repugnant
and that we should not overfeed the utility monster, without giving
up on utility calculus entirely, seems to 
require drawing some boundaries between ``normal'' and 
``extreme'' cases.   The lack of obvious bright
lines or consensus on principles that could establish
such boundaries somewhat undermines confidence in the 
applicability of utilitarian reasoning, even in 
more realistic policy decisions where it 
seems in line with common sense.

Nonetheless, quasi-utilitarian arguments do seem to 
capture many people's intuitions
even on some existential questions.  
Many agree both that the extinction of
human life would be a much
greater disaster than the death of $99\% $ of the 
human population and that one can base a good argument for
this on a comparison of the numbers of all future lives 
lost that would otherwise be lived in each case \cite{parfit1984reasons,kent2004critical}. 
Many also accept the corollary that we should devote
significant resources to accelerating human migration from Earth,
to mitigate the extinction risk from a planetary
catastrophe.   
Some, though probably far fewer, also believe we have a moral imperative to colonize
as much of the cosmos as possible as fast as possible, even at the cost
of great present and future sacrifices, in order
to maximize the number of lives well lived in our
future light cone. 

Further problems arise if we try to give 
utility calculus a scientific foundation.   
Must an agent be conscious in order to 
have utility?   Are there degrees or different qualities of consciousness, and
are they ethically relevant?  Could there be a meaningful sense
in which two agents have the same experiences but
one has a ``thicker'' consciousness than the other, and
hence more utility, and hence more ethical weight?   
Is the utility of experience reducible
to the utilities of individual qualia?   There is no consensus
on whether these questions are meaningful, nor on
the most plausible answers if they are.  

In short, most of us are in an uncomfortable position intellectually. 
We are embarrassed that we do not really know what utility is 
or how to compare utilities of different agents.  
Even in situations where we accept a given model of utility, 
we use utilitarian reasoning inconsistently. 
Sometimes it feels reasonable; sometimes it feels
unpersuasive or clearly wrong.  Most of us struggle to 
articulate where and why we draw boundaries.
We may be clear enough about 
some cases on one side or the other to argue for
some boundary criteria, but these tend to be very incomplete,
leaving many cases undecided.  
We may perhaps accept health care rationing \cite{rawlins2004national}
based on utility guesstimates, such as that a year of 
life for two typical individuals is of comparable
value, perhaps with some quality of life adjustment
factor that might typically range from $10^{-1}$ or so
to $1$.   But if we do, our 
acceptance tends to be pragmatic -- some politically defensible
policy is needed, and there seem few alternative candidates.
On more general policy questions, where costs and benefits
apply to different subsets, we flounder.  

These problems deepen still further when we consider the 
possibilities of transhumanism, the development of 
general purpose artificial intelligence at human
or superhuman level, and our options for steering the evolution
and spread of life through the cosmos. 
We are very poorly equipped to
address these increasingly pressing questions. 
A dappled view of future population ethics, 
combining the most valuable insights from many perspectives, may be the
best we can hope for. 
One way to improve our understanding of the possibilities
is to consider thought experiments 
that allow us to test out and illuminate utilitarian 
and other ideas and to identify points where different plausible
intuitions produce very different conclusions. 

\subsection{Whole brain emulation: possibilities and caveats}

Various technological routes that might lead to the 
development of general purpose human and superhuman level
artificial intelligence are being discussed and pursued:
see for example Ref. \cite{bostrom2017superintelligence} for a survey
of possibilities.   
Expert opinion differs widely on which of these is most
likely to succeed, and on plausible timescales.
There are several reasons for this.   
Humanity faces many challenges, and it is not 
certain that technologically advanced
economies will survive long enough to develop
general purpose human level AI.\footnote{One 
reason for uncertainty is that, while current and foreseeable 
technological developments may fall far short of attaining general 
purpose human level AI, they already have the potential to transform societies and 
global power dynamics, in ways we seem ill-equipped to handle.} 
Even if we assume that relatively stable societies continue
to support scientific and technological 
research and development, foresight in these areas
is notoriously challenging.   The history of  
controlled nuclear fusion research is one of many cautionary
examples.   
Finally, in apparent contrast to controlled nuclear fusion, 
whether human level AI is achievable by any given
technology depends on unanswered fundamental scientific questions.    
We do not know in detail how human brains work: we do not know, for
example, whether a description at the level of neurons and
synapses is sufficient to explain human brain function.
We do not have a scientific understanding of consciousness, 
or even confidence that it is possible to find one, 
and we do not know whether such an understanding 
is necessary in order to fully describe and understand 
human brain function.\footnote{Or, to put it neutrally, there
are deep disagreements on the problems of consciousness 
and their implications, with no consensus view.}

One of the strategies proposed as a serious
contender is whole (human) brain emulation. 
The proposal is that sufficiently advanced scanning
technologies can identify all the functionally 
relevant features and processes of a given human's
brain at a given point in time, and that these
can be emulated by software running on sufficiently
fast and large computers.  Surveys and tentative
assessments of the relevant technologies can be
found in Refs. \cite{sandberg2008whole,sandberg2013feasibility,bostrom2017superintelligence};
recent technical developments are reported in Ref. \cite{jordan2018extremely}.  
An attempt at a sketch of a plausible world economically dominated
by such human emulations, with justificatory arguments
based on fundamental physical and economic principles, 
was offered in Ref. \cite{hanson2016age}.  
 
Discussion of whole brain emulation involves many 
uncertainties, as the authors cited above acknowledge.   
Whole brain
emulation relies on knowing what the functionally
relevant features and processes in the human brain
are.   In particular, it relies on the hypothesis
that there {\it are} finitely many functionally
relevant features and processes that suffice to
describe all of human behaviour.  
Even if this is true, it does not follow (unless
one takes a functionalist view of consciousness)
that a human brain emulation on a computer is 
necessarily conscious.   We have no relevant
firm facts about consciousness, and one can
imagine theories of consciousness in which
the existence or type of a consciousness
depends on its physical substrate 
as well as its functional organisation.   

Even if human brain emulations are conscious, 
it does not necessarily follow that an emulation of a given
person's brain can or should necessarily be identified 
with that person, in the same sense that we identify
our brain and mind states now and one second later
as belonging to the same person.     
Personal identity is poorly understood.
It may perhaps be that Parfit \cite{parfit1984reasons}, whose arguments we
consider in more detail below, is correct that 
what we conventionally consider to be personal identity ultimately
has no stronger justification than psychological continuity.   On this view, we should consider our
future selves as being essentially the same person 
as our present selves only because (and only insofar as)   
our future thoughts will be highly and intricately correlated with our 
present thoughts.   But one can imagine that there are further special
facts, whose nature we cannot presently imagine, but which will turn
out to imply a different definition that seems relevant, perhaps
even persuasive, to many.  

The example of parenthood is an interesting partial analogy, 
which illustrates the problems in reaching a definitive conclusion. 
The analogy is complicated in various ways, including the fact that the birth mother of a 
child has always been identifiable, while the biological father
has not been: indeed, even the concept of biological father has
not been available until relatively recently.    
Because the conceptual shift in the light of scientific discoveries
has been larger, it is helpful to focus here specifically on fatherhood.  
A pre-modern analysis of fatherhood, in the case of a 
child of a mother known to have had more than one recent sexual
partner, might reasonably have suggested that in some societies there was no 
meaningful distinction between {\it being} the child's father
and {\it behaving as and being treated as} the child's father.   If the mother
accepted a recent male sexual partner as the father, and he accepted the role
of father, then he was the father.   There were, it might reasonably have
been argued, no special further facts that could definitively decide
otherwise (although strong physical resemblance to one or another
sexual partner might perhaps have been taken as indicative evidence).   

Now, I do not want to suggest this conclusion was wrong.  
Indeed, some would argue that social changes and medical developments
have led to a more widespread understanding that parenthood {\it should} 
be understood as the adoption of a role, rather than a biological
relationship.  I certainly do not want to argue against this view. 
But I do want to note the salient point: we have learned, after the discovery of DNA and its
role in genetics, that there {\it are} further facts that some
individuals and some societies regard as relevant in
some circumstances.  These further facts allow us to define and
identify a child's biological father
(and, in the rarer cases where there might otherwise be ambiguity,
biological mother).
And these are not only of interest to geneticists:
the status of biological parent has legal and ethical significance in
many societies, and also clearly has personal significance for many
(though not all) parents and
children.   In short, although individual views on the relevance
of biological parenthood vary, the discovery of DNA genetics has changed our collective 
understanding and treatment of parenthood.    

It is completely unclear what form a fundamental scientific theory of
consciousness might take, if indeed there is one, and it is also 
unclear whether and how it might shed any new light on personal
identity.   But the example of parenthood illustrates that it is
not ridiculous to imagine the
possibility.   Parfit \cite{parfit1984reasons} may be right that we presently 
have no good reason to reject a definition of personal identity
based on psychological continuity, because we presently have no 
relevant further facts available.
It may also be right to say that no further facts could possibly  
logically refute this definition: the concept of personal
identity may well be subjective and elastic enough to leave room for 
a variety of definitions, whatever relevant
scientific facts we may uncover in future.   
But we can nonetheless imagine discoveries that seem persuasively
relevant to many.   These might (or might not) suggest reasonable 
definitions that imply, for example, that
a given person can only be identified with one future successor,
even when two or more successors are  functionally identical, 
or that typical emulations of a human brain on a standard computer
do not share a personal identity of the brain's owner, even when
the emulations appear functionally equivalent to the original.  
These give further reasons for being cautious in predicting the future
development of whole brain emulation technology and its 
effect on the world.   

\subsection{Whole brain emulation as a motivating hypothesis} 

Hypothetical whole brain
emulation technologies raise interesting ethical questions 
about the ethical weight attachable to emulations.
These questions are interesting whether or not human brain emulation is the first 
form of human level artificial intelligence we
achieve, since it will presumably be achieved (perhaps
by superhuman level AI) at some point if achievable, and the
questions will apply directly at that point if humans
are still around.    
They are also of much broader relevance.   
For example, whatever forms of artificial intelligence 
we develop, they can presumably be replicated, and
if we attach any ethical weight to them then similar
questions apply.   Also, as we note later, the questions
shed an interesting light on the weight attached to
diversity in present day population ethics applied to other species 
and in principle (though with many important caveats) to humans.   

So, although we take very seriously the caveats about human brain
emulation, we are motivated to neglect them for the purposes of
the following discussion.   Perhaps human brain emulation will work;
perhaps there are no further relevant facts about personal identity
beyond psychological continuity; even if there are, perhaps they
will not be regarded as decisive by many.  In any case, the questions
will very likely be relevant when considering other
forms of AI, and are already relevant when considering existing
forms of life.   To keep the discussion here as simple as possible
and focus on our essential point, we will assume that human
brain emulations function, are conscious, and maintain 
personal identity, even when there are multiple emulations
of a single brain.   Alternative hypotheses are also plausible
and interesting.  However, the simplest alternative hypotheses lead
to obvious conclusions -- for example, if human brain emulations
are not conscious, then they presumably have no ethical weight. 
More complex ones deserve separate motivation and discussion;
presently lacking the former, we leave the latter for future work. 

\section{A thought experiment} 

The main point of this article is to propose a thought experiment which could become a 
real dilemma for programmers, researchers and policy
makers in the not necessarily far distant future. 

\subsection{Human emulations}

Suppose, for the sake of argument, that something 
roughly like an individual
human life can be lived through software running on a 
computer.\footnote{I do not mean to suggest this is evidently true.  
I think it's quite possible we could yet find facts about the world that 
make it seem implausible.  It's also possible that it will
always continue be a plausible position that remains open to reasonable doubt.}
Here ``computer'' means
something like, although more advanced than,
the things we conventionally call computers at present. 
Very roughly speaking, this means a device assembled from raw
materials into a specific architecture, by mechanical means, 
in order to process classical or quantum information, in a way 
that can be programmed by a choice of inputs.\footnote{We want the
term ``computer'' to exclude biological humans here, even though
human children can be said to fit most of these definitions.
We also want to avoid considering skeuomorphic boundary cases,
not because these are implausible or uninteresting, but because
they complicate the discussion in a way that obscures the 
key points.    
We are interested in discussing human emulations whose
physical substrate is radically different from our own.}
The ``software'' could,
for the sake of the argument, perhaps be obtained 
from scanning the brain of a biological human with
sufficient resolution to allow every relevant detail of 
brain function to be transferred to and simulated on
the computer.\footnote{Scenarios in which such emulations dominate 
the world population, at least temporarily, are discussed in Ref. \cite{hanson2016age}.}   We will suppose 
without further discussion that we have reasons that seem to us compelling enough
for treating such simulated lives as comparably valuable to an
ordinary human life.\footnote{
It is not necessary for the discussion here to assume 
specific reasons.  Thoughtful people who agree 
that there may well be good reasons disagree
on what the reasons may be.  For example, 
some think it likely that simulated
humans are conscious in essentially the same way
as their biological counterparts, and feel happiness,
pleasure, pain and other qualia with just the same qualities
in just the same quantities.    
Others believe or suspect that this language is misleading or ill-defined, 
and would say that what ultimately matters when considering
whether an entity is valuable is the functionality 
of its information processing.}

Suppose also that this has become practical:
we can create a human emulation relatively cheaply. 
If we can emulate a human individual once in software, it seems
very plausible that we can easily create many
emulations of the same human.   Let us assume
this too, for the sake of argument.\footnote{Again, I do 
not mean to suggest this is evidently true.   For example,
it could turn out to be a fact of the matter that human identity
is uncopiable, just as unknown quantum states are
\cite{wootters1982single,dieks1982communication}, 
and possibly even for that very reason.} 

Most likely, we will then find it easier to 
create $N$ emulations of the same human
than to create emulations of $N$ different humans.
For example, if brain scanning is the relevant technology,
we would only need one brain scan rather than $N$. 
It seems very plausible that brain scanning would require
significantly more resources than copying the software that
results from brain scanning.
If so, scanning one brain and then making $N$ copies of a single
emulation would require fewer resources than scanning $N$ different
brains and making emulations of each.   
It might also be possible to improve the 
efficiency of a human emulation significantly, with some analysis
and optimization work, so that it needs significantly
fewer gates and/or less energy, and perhaps runs 
significantly faster.   Again, this is work that 
only needs to be done once if we create $N$ emulations
of the same human.\footnote{Even if the work itself
is quite resource-intensive, it becomes worthwhile if 
$N$ is sufficiently large.} 
So, if emulating the same human $N$ times seems to us to be just
as worthwhile as emulating $N$ different humans,
efficiency would motivate us to prefer the former.  

\subsection{Further assumptions}

To make our thought experiment as clear as possible, 
we need to make further simplifying assumptions.  
We assume that our emulations are in a self-contained
and effectively isolated computer.   They have no
effect on the world outside the computer; in particular
they are not observed by outsiders.   We are creating
these emulations purely because we believe that a typical
human life is an intrinsic good. 

It may reasonably be argued that a single human life
in isolation would be a miserable existence, and so
{\it not} an intrinsic good.   
To counter this, we assume that along with the emulation
we simulate an environment rich enough that it seems
to the emulated human that they are interacting with
many other humans in an interesting 
world.

This may perhaps still seem unrealistic.
One might query whether it is possible to simulate the experience of interacting
with other humans without actually simulating the other
humans.   Even if it is possible, it might be argued that there is no 
value in an emulated human life so deceived about their true 
situation, whatever the emulation's beliefs on the matter.    We could counter such arguments
by considering instead emulations of a large community of 
humans interacting with one another and with a rich environment, running over many
generations if desired.   Our options then 
would be to emulate exactly the same community and
environment $N$ times, or to emulate $N$ different
communities in different environments.   
For definiteness, and 
to keep the language simple, we will speak about 
emulating one human $N$ times or $N$ humans below.
However, the discussion is meant
to generalize: ``emulated human life'' may be read
as shorthand for however large and however long-lasting
an emulated community
and environment are required to assure us that the
emulation is an intrinsic good.  

It is important to note that, by assumption, our
emulations are deterministic rather than probabilistic.
Because the emulations are self-contained and isolated,
there are no random influences from the outside world
that cause them to run differently.   Because we 
replicate not only the human (or community of humans)
but also their modelled environments precisely, 
they all interact identically with their modelled
environments.   If the emulations require random
inputs -- for example, to emulate the humans carrying
out quantum experiments within their emulated environments --
then we assume that these are drawn from a pre-generated
random string, with the same string being used for each
emulation.   So, in each emulation, the emulated humans
see the same random outcome of any experiment or observation
that involves randomness.   

\subsection{The experiment}

Suppose for some reason that we are restricted
to four choices:

\begin{itemize}

\item Emulating the same life of the same human, Alice, 
$N$ times, in parallel on different computers, so that
the lives all run simultaneously.

\item Emulating the same life of the same human, Alice,
$N$ times, in series on the same computer \footnote{Here 
and below different computers could also be used: the precise physical instantiation
is not relevant to our arguments.}, so that the 
lives run sequentially.

\item Emulating $N$ different lives of $N$ different
humans, Alice, Bob, $\ldots$, in parallel on different computers, so that
the lives all run simultaneously.

\item Emulating $N$ different lives of $N$ different
humans, Alice, Bob, $\ldots$, in series on the same computer, so that the 
lives run sequentially.

\end{itemize}

Suppose also that in all cases the relevant humans cannot 
continue living in the world outside the emulation.   
Alice, Bob, and the others are $21$ year old adults who are terminally ill
today, with a disease that has ravaged their bodies but has left their
brains intact.   We just have time to scan their brains before they die.  
We can then emulate them in software, and, so to speak, 
continue their lives as emulations from age $21$ onwards.   
Suppose further that all the candidates are equally admirable
and will have equally valuable emulated lives, by any reasonable criteria. 

In each case, the emulated lives are unobserved and have no effect on the rest
of the world, which continues just as it otherwise would were no
emulations run at all.   The only reason for thinking any of 
these courses of action is worthwhile is because the lives have intrinsic value. 

Are any of these choices preferable, and if so why? 

\subsection{Temporal considerations}

Does it matter whether the lives are run in parallel or series?
Is it more worthwhile to create a present life than to create
a future life, assuming that in either case the lives are 
self-contained?    

Evidently, the two cases would be different if there were some
risk that some of the future lives might in fact not be created.
Emulating $N$ lives in series takes longer than $N$ lives in
parallel.   For large $N$, there could be a significant risk of a 
catastrophe destroying us and our technology before a series of $N$  
lives ends.  For the sake of the argument, let us assume there is no such
risk: we are certain that all the emulations will complete, whichever
choice we make.  

Another possible reason for preferring present human 
lives to future human lives is that we may expect to have more in common ---
more common memories, attitudes, overlaps and continuities 
of thought -- with contemporaries than with descendents.
As Parfit has argued \cite{parfit1984reasons}, this may also be a reason to give more
weight to the interests of our present selves than our
future selves. 
However, it does not apply in our experiment.   
Alice, Bob, and the others are all contemporaries of ours. 
Their emulations will begin with the same memories, attitudes and
thoughts, whether they are run immediately or in the distant
future.  Since they do not interact with the world outside
the emulation, they will continue in the same way, whenever
the emulation is run.    

Discounting these reasons, it seems hard to find any justification
for preferring either parallel or series emulations.    They have
the same total intrinsic value, and it is hard to see why it should 
matter whether that value is realised during one emulated lifetime
or in the course of $N$ sequential emulated lifetimes,
ceteris paribus.\footnote{Thinking relativistically, can it 
really matter whether two emulated lives are spacelike 
separated, with life $A$ just to one side of the future light cone
of life $B$, or timelike separated, with life $A$ moved just to 
the other side of the future light cone?} 

\subsection{Replication considerations}

Does it matter whether $N$ different lives are emulated, or 
the same life is emulated $N$ times?   
This seems trickier to adjudicate.  

One way to think about the problem is that each emulated life
has a value $V$, which by assumption is positive.   
Emulating $N$ different people thus has value $NV$. 
But emulating Alice $N$ times independently
also has value $NV$.   These valuations give no reason
to prefer one option over the other.   On this view, if, as seems
plausible, it costs fewer resources to emulate Alice
$N$ times, then we should prefer to do that.   

Opposing this are intuitions that have a weak and strong form. 
The weak form -- {\it replication inferiority} --
is that emulating Alice $N$ times in sequence seems 
less valuable than emulating $N$ different people.
The strong form -- {\it replication futility} -- is that emulating Alice
$N$ times is no more valuable than emulating her once. 

Neither of these intuitions can be justified in our experiment
on the grounds that any of Alice's emulations know about the
others and find this distressing.  By assumption, each emulation is independent,
and has no memory, awareness or foreshadowing of the others.    

\subsubsection{Replication futility}

Rather, the intuition behind replication futility is that, having written Alice's life 
into the universe once, we add nothing to 
the total value of the universe (integrated over
space and time) by doing so again.      
In particular, we add no value {\it for Alice} by 
repeating her emulation, even though each emulation of Alice has the sense
of appreciating their life.  

This intuition can be held even if one believes that repeated
emulations are distinct lives, in some meaningful sense that may
some day be justified by fundamental physics.  
Suppose one believes that there are facts of the matter about
experiences in time and space, and that it is a fact of the matter
that the first emulation has experiences, and the
other emulations have the same experiences, at different times
or in different places.   One can still consistently believe that repeating
these experiences is valueless.

There is another possible view that more strongly implies replication futility.
This is that there is no meaningful sense in which repeated emulations
(whether in time or space) are distinct lives.   The strongest form
of this view is that there is a fact of the matter about experiences
in time and space, and that it is a fact of the matter that 
all the emulations correspond to only one set of experiences, 
and so effectively define only one life.   While this might
at first blush seem a peculiar idea, it is consistent with
some views about the relation between material physics 
and conscious experience.  If consciousness is defined 
only by the functionality of information processing, 
no facts about its physical instantiation -- including
whether or not it is multiply instantiated in space or time -- should 
affect the nature of the conscious experience.

Indeed, it is hard even to define a sharp and general
distinction between single and multiple instantiations.
Does a computer running an algorithm constitute a 
single instantiation?   Or can it equally well be 
seen as several parallel instantiations, by subdividing each
wire of the circuit?   One can ask similar questions of
other macroscopic information processing systems, 
including our own brains.     
This gives some way to build an intuition that
replicated emulations in space, even if well separated,
might correspond to a single set of experiences.  
It may seem less intuitive that replicated emulations
over time also could.   Indeed, there could conceivably be 
a fundamental distinction between the two cases.\footnote{
Again, thinking relativistically gives some reason to query
how plausible it is that the cases really are separate.
One would have to suppose that $N$ spacelike separated 
instantiations correspond to a single set of experiences,
which splits into multiple sets of experiences if any
instantiation crosses the future light cone of any others,
and recombines again if it crosses back.}
Still, if we accept the principle that the functionality
of information processing is key, and hence that replicated 
emulations in space could correspond to a single set of
experiences, it is hard to see a presently compelling reason why
replications in time could not also.  Our perception of a flow of time 
is even less well understood and more fundamentally puzzling than
our consciousness: we have no firm facts on which to base a 
refutation.    

\subsubsection{Replication inferiority}

One intuition that might lie behind replication inferiority is that, even if we 
do perhaps add some value to the universe by replicating Alice's
life, the first replication adds less than the original, the
second less than the first, and so on. 
This may seem to follow from the general assumption that utility
functions should be convex.  However, the standard reasons for
convexity do not directly apply.  We are less grateful for the eleventh orange
than the first.   It is not so clear whether we should also value the eleventh Alice
less than the first.   We become jaded with oranges, 
but cannot become jaded with Alices in the same way, since we 
do no interact with any of them.   Nor can Alice become jaded with
lives in the same way, since each emulation has no memory of the
others.   Nonetheless I find it hard to escape the intuition that 
a universe with a billion independent identical emulations of Alice is 
less interesting and less good a thing to have created than a 
universe with a billion different individuals emulated.
I find it hard too to escape the intuition that, while I might
be happy for the promise of an extra year of life in a future emulation, 
my gratitude if promised also that the same year would be replicated precisely in 
independent emulations would be at best muted.    

Another intuition that might also lie behind replication inferiority
is a sense of cosmic justice.   Even if we think a replicated emulation
benefits Alice just as much as the original, we might still think it 
unjust, or at least less than optimally beneficient, to 
give Alice the benefit of $N$ lives 
and no one else the benefit of any, when we have the alternative 
of giving $N$ people one each.\footnote{Related points
are made in critiques of attempts
to use decision theory to explain the relevance
of probability in many-worlds versions of quantum theory; see in
particular Refs. \cite{price2010decisions,kent2010one}.}

\section{Parfit on replicated selves}

In ``Reasons and Persons'', Parfit considers 
thought experiments involving duplicated selves.
One of these is a hypothetical brain splitting, 
in which an individual's two brain hemispheres
are transplanted into two separate donor bodies. 
The hypothesis, for the sake of discussion, is that
this can be done so that not only do both operations
produce fully functioning humans, but each of them
has psychological continuity with the original individual. 
In another thought experiment, a Replicator machine scans an individual's
brain and body at location $A$, destroying them, but sending data allowing
the individual to be reconstructed at two other locations, $B$ and
$C$. 

The brain splitting hypothesis involves a number of assumptions --
that hemispheres can have separate consciousnesses, that these
can survive transplantation, and that each consciousness can
fully psychologically represent the individual who 
once had a unified consciousness arising from the two
connected hemispheres.   Parfit cites some evidence
for the first of these, and the second seems conceivable in
principle.  However, the third seems unlikely to be precisely
correct.   All in all, the combined hypotheses raise many doubts
and complications.    Brain splitting
involves messy biological and psychological
questions that seem very unlikely to have the idealized answers
hypothesized.    

The second thought experiment also involves questionable assumptions, of course.   
Human scanning and reconstruction may never be viable;
scanning of classical information about the human would
have to suffice for it to be possible to reconstruct
more than one copy, and we do not know whether classical
information suffices; we do not know whether Replicas 
can or would be conscious, or even functional, and if
so under what conditions.    Still, it seems a cleaner
thought experiment, in that the assumptions involve 
the fundamental relations between matter, consciousness
and identity, and it seems at least possible that these
{\it could} take forms that make the experiment
viable in principle.   

Another significant difference between Parfit's hypothetical
duplicates and our hypothetical emulations is that Parfit
imagines his duplicates will continue living in the same world,
able to interact with one another and with friends and 
lovers of the individual from whom they originated. 
As Parfit comments, this complicates any measure of
their value.   It could be unpleasantly uncanny to 
live in a world with a duplicate, and upsetting to 
find your friend or lover duplicated, and these outcomes would
subtract value.  On the other hand, duplicates could possibly
satisfy two deeply held but incompatible ambitions of their
originator, and this could add value. 

In considering value, Parfit's focus is on the value to 
the original individual.   Should he consider his 
destruction and recreation in duplicate as equivalent
to -- roughly as bad as -- death?   One possible 
reason for doing so is 
that he cannot justify identifying himself with one 
specific successor, and it seems incompatible with 
most intuitions about personal identity to 
identify oneself with two successors.  The remaining
option, identifying oneself with no successor, appears 
then to fit both our understanding of duplication and
our understanding of death.  

Parfit rejects this conclusion.  
He argues that psychological continuity
is all that matters; there is no special further fact
about personal identity.   Having two future duplicates
who have psychological continuity with your present
self is not the same as having no future individual
who has psychological continuity with your present self:
``Two does not equal zero.''   (Ref. \cite{parfit1984reasons}, p.278.)

But does ``Two equal one.''?   On this, Parfit is 
less clear.    
He notes that a relevant consideration in 
comparing his own duplication and his survival in the ordinary sense is 
\begin{quotation}
``the fact that these two
lives, taken together, would be twice as long as the rest of mine.'' 
(op. cit., p. 264) 
\end{quotation}
Parfit allows that this may indeed be a reason for preferring duplication
to survival:  
\begin{quotation}
``Instead of regarding division as being somewhat worse than ordinary survival, I might regard it as being better. The
simplest reason would be the one just given: the doubling of the years to be lived. I might have more particular
reasons.'' (op. cit., p. 264) 
\end{quotation} 
However, after discussion, he concludes: 
\begin{quotation}
``The best description is that I shall be neither resulting person. But this does not imply that I should regard division as
nearly as bad as death. As I argued, I should regard it as about as good as ordinary survival. For some people, it would
be slightly better; for others, it would be slightly worse. Since I cannot be one and the same person as the two resulting
people, but my relation to each of these people contains what fundamentally matters in ordinary survival, the case
shows that identity is not what matters.''  (op. cit., p.279.)
\end{quotation}

Duplication may be better or worse than survival in Parfit's
scenario because of the complicating factors already mentioned --
there are potential pros and cons to having duplicates interacting in 
the same world.   It may also be preferable because of a doubling of 
the years to be lived.   But the net result of all these considerations, in Parfit's view,
is that it should be about as good as survival. 
Now it is true that the phrases ``For some people'' and ``for others'' do not
logically cover all the options.  Logically, the text is consistent
with the possibility that Parfit regards duplication as about as good
as survival, some people regard it as slightly better, others as
slightly worse, but others still as greatly better or greatly worse. 
That said, a natural reading is that Parfit
thinks that for most or all people duplication is likely to be at best slightly
better than survival and at worst slightly worse.   
If so, then none of the relevant factors, including the doubling of
years to be lived, can generally carry much weight.    

Parfit does not explain {\it why}
doubling the years lived carries little weight for him. 
This is a rather surprising omission, given that it is explicitly
considered as a factor.   However, Parfit's central concern is to argue that 
duplication is a form of survival rather than a form of death. 
It may be that he did not think it necessary to  
articulate a precise 
position on the relative values of duplication and survival.   
The quotations above, though, suggest an inclination towards 
the view that replication futility applies, at least approximately,
in our thought experiment.  
This would imply that replication inferiority
certainly applies.  
  
\section{Noisy emulations} 

We now return to our thought experiment involving independent 
self-contained emulations, isolated from each other and the
rest of the world.   Suppose we have concluded that
either replication inferiority or replication futility applies
in the experiment.  
And suppose now we vary the experiment slightly, by
assuming that the emulations are not quite perfect
and not quite identical.    Perhaps once or twice in
an emulated life, a pixel of the visual field is
briefly a slightly different shade, or an emotion
is very subtly modulated, or something of the sort,
because of a brief and minor glitch in the emulation.
Suppose too that these brief alterations have no 
longer term consequences for the emulation.  

These differences seem too minor to significantly affect
our value judgements.   If a duplicated emulation has no more value 
than a single emulation, then a duplicated emulation with small
glitches in one copy surely can at most have very little more value. 
Our own life would have almost exactly the same value 
if it were slightly and briefly modulated in this way;
not only that, but the modulated life would seem almost precisely
equivalent to ours.   Intuitively we imagine a sort of continuity principle applying
to the value of lives, along with some bounds on the derivatives. 
And we imagine the same sort of principle applying to $N$ emulations:
$N$ perfect emulations should have almost 
precisely the same value as $N$ slightly and briefly modulated
emulations.  This should be true even though the original $N$ emulations
are precisely identical and none of the modulated $N$ emulations need
be precisely identical (because their slight and brief modulations are
all different).   It follows that, if replication futility holds
for identical emulations, it should also hold to very good approximation   
for near-identical emulations.   Similarly, if replication 
inferiority holds for identical emulations, it should also hold for
emulations that are not identical but are sufficiently close
to being identical.    

We can press a little further.    Even if the modulations
persist throughout an emulated life, the same conclusions should
hold, so long as the modulations are sufficiently small. 
It should not matter that much if an emulation has slightly
modulated experiences of shades of the colour red throughout
his life, so long as these have no other significant experiential 
consequences.    

\section{Valuing variety?} 

It seems good to be as clear as possible on how we value lives, before we unwittingly set in motion 
well-intentioned research that could end with an artificial
intelligence trying to optimize the number of emulated 
human lives per unit $4$-volume in our future light cone.
Unless we want it to find the most easily emulated human
life and run it everywhere, we should give it different
value rules.  Suppose we have concluded that replication
futility applies in our thought experiment.
We can then tell it (say) that identical lives do
not have independent value.
However, if we tell it only this, its next attempt at 
optimization is likely to involve small and easily 
implemented variations on a single life.   
To avoid that, we need to tell it that, as we have argued,
$N$ sufficiently similar lives have at most
little more value than $N$ identical lives.     

It will then, of course, ask for definitions of  
``sufficiently similar'' and ``little more value'' and for more general rules.
Can we give it a general calculable measure $d(L, L')$ of distance
between lives $L$ and $L'$, and some continuous function 
$\delta: {\mathbb R} \rightarrow {\mathbb R}$ with $\delta (0) = 0$, 
such that ``$L$ and $L'$ are
sufficiently similar to be of roughly equivalent value'' translates as 
\begin{equation}
d(L,L') < \epsilon \implies | V(L) - V(L') | < \delta( \epsilon ) \, ?     
\end{equation}
And then can we give it a value function on $N$ lives,
\begin{equation}
V( L_1 , \ldots , L_N ) 
\end{equation} 
and bounds on $V(L_1 , \ldots , L_N )$ as functions of (say)
$V(L_1 )$ (or $L_1$ ) and of the distances $d(L_i , L_j )$?  

By this point we realise that we 
have introduced the thin edge of a potentially very large
and very contentious wedge.    
For example, we have allowed the logical possibility that there
are sets $\{ L_1 , \ldots , L_N \}$ of {\it existing} human lives that 
have sufficient similarities (pairwise or collectively) that
some defensible value measures $V$ might give them significantly less
value than $N$ lives randomly chosen from the Earth's present
human population.    This possibility need not be realised
in practice for any value measure we regard as defensible, of course. 
We can politely, plausibly and perhaps wisely take the
view that any sensible similarity measure $d$ necessarily has the
property that $d(L, L') = \infty$, for all practical purposes, 
for any pair of distinct presently existing human lives $L,L'$. 
We might further hope that plausible
axioms can be found that justify this view.  
\vfill\eject
However, other views are possible.  
As the song goes: 
\begin{quotation}
$\ldots$ the people in the houses \\
All went to the university, \\
Where they were put in boxes \\
And they came out all the same, \\
And there's doctors and lawyers, \\
And business executives, \\
And they're all made out of ticky tacky \\
And they all look just the same.  \cite{reynolds1963little}
\end{quotation}
One gets the sense that Reynolds and Ball might have valued $N$ of 
these people less than some other sets of $N$ 
humans, not \footnote{Or not mainly.} because of their professions
and lifestyles but because of their perceived similarity.\footnote{I like
the song.    Still, anyone entirely comfortable 
with replication futility applying axiomatically to 
doctors, lawyers and business executives 
ought also to consider the many other groups -- including
their own -- to whom it might equally be applied.} 
 
Even if we are comfortable that present day humans are dissimilar
enough for replication inferiority considerations to be irrelevant,
are we so sure this applies to all denizens of Earth?  
Is the life of one worker ant really significantly dissimilar from
that of another of the same species, in a similar environment?  
What about a cow or sheep?   Or a flatworm or an amoeba?     

And are these considerations more pressing when we think about 
filling our future light cone with human and post-human life?    
Even if the human species were to continue with traditional
sexual reproduction, and continue with phenotypes and genotypes
broadly in the spectrum of ours, in environments broadly similar
to ours, questions of replication inferiority may not ultimately
be avoidable.  
If the ``space of possible lives'' is modellable by a compact 
set, and the similarity function $d$ by a metric, 
then for any $\epsilon > 0$ there must be {\it some} 
population size $N$ such that at least one pair $L,L'$ necessarily have
$d(L , L') < \epsilon$.  
In other words, there must be some population size such
that at least one pair have sufficiently similar lives for replication inferiority considerations
to come into play.   Perhaps we may end up concluding that 
a compact set with metric is not a good model.   Or we might
conclude that, for 
sensible measures $d$, the relevant $N$ is much larger than the number
of humans that can populate our future light cone between now and
the end of the universe.   But these are not {\it a priori} obvious 
conclusions.   

In any case, in the long run,
our descendents are mostly unlikely to have similar
genotypes, phenotypes or psyches to ours. 
They may well include genetically engineered humans, 
humans in varying types of symbiosis with artificial intelligences,
emulated humans, and skeuomorphic hybrids of several of these. 
These technologies will greatly stretch the possibilities,
but they may plausibly also make intelligent life in the 
cosmos (at least that of terrestrial origin) much less diverse
\cite{hanson2016age}.   The value of (near-)replication may well be 
a clear and pressing concern for them.   In so far as we 
can influence this future world by present decisions, it 
is a question we need to take seriously.

\section{Conclusions}

Decisions we are beginning to make now about the future
of life involve hard questions of principle.
One of those is what we mean by saying that lives
are very similar, and how to value sets of lives that are. 
We have tried here to clarify the questions somewhat, 
and to suggest ways of categorising possible types of answer.
We do not have fully satisfactory answers; so far as we are 
aware, no one presently claims to. 
These problems deserve further consideration. 

Our lines of thought give new perspectives on existing
problems.   For example, if (pace Tolstoy \cite{tolstoy1966anna}), 
the $M$ lives in the Repugnant Conclusion scenario 
that are barely worth living are -- necessarily, as a 
consequence of their very low utility --
also very similar in their drabness, then replication
inferiority or futility offer ways to avoid the Conclusion.    
However replication considerations, if anything, add to the
case for feeding Nozick's Utility Monster: we have similarities
to one another, whereas the Monster is unique. 

Some may perhaps wish to align the idea of suboptimal utility of 
replication with political programmes aimed at altering the 
characteristics of current human societies. 
We strongly urge caution.  As noted earlier, it is plausible that
any sensible distance measure obeys $d(L,L') = \infty$
for any pair of human lives in the near future.   Measures with other properties
are also logically arguable but such arguments risk being no more than lightly 
coded attempts to dehumanize some disfavoured 
groups.\footnote{Again, no one should assume that 
replication inferiority considerations self-evidently 
support a particular party line.   If applicable at all, they could
potentially
be applied to plutocrats and peasants, babies and bankers, academics
and assembly line workers, activists and the aged.}

{\bf Acknowledgments} \qquad 
This work was partially supported by an FQXi grant and by
Perimeter Institute for Theoretical Physics. Research at Perimeter
Institute is supported by the Government of Canada through Industry
Canada and by the Province of Ontario through the Ministry of Research
and Innovation.   
	
\bibliographystyle{unsrtnat}
\bibliography{doppelgangers}{}

\begin{thebibliography}{16}
\providecommand{\natexlab}[1]{#1}
\providecommand{\url}[1]{\texttt{#1}}
\expandafter\ifx\csname urlstyle\endcsname\relax
  \providecommand{\doi}[1]{doi: #1}\else
  \providecommand{\doi}{doi: \begingroup \urlstyle{rm}\Url}\fi

\bibitem[Woollard and Howard-Snyder(2016)]{sep-doing-allowing}
Fiona Woollard and Frances Howard-Snyder.
\newblock Doing vs. allowing harm.
\newblock In Edward~N. Zalta, editor, \emph{The Stanford Encyclopedia of
  Philosophy}. Metaphysics Research Lab, Stanford University, {W}inter 2016
  edition, 2016.

\bibitem[Parfit(1984)]{parfit1984reasons}
Derek Parfit.
\newblock \emph{Reasons and {P}ersons}.
\newblock OUP Oxford, 1984.

\bibitem[Nozick(1974)]{nozick1974anarchy}
Robert Nozick.
\newblock \emph{Anarchy, {S}tate, and {U}topia}, volume 5038.
\newblock New York: Basic Books, 1974.

\bibitem[Kent(2004)]{kent2004critical}
Adrian Kent.
\newblock A critical look at risk assessments for global catastrophes.
\newblock \emph{Risk Analysis}, 24\penalty0 (1):\penalty0 157--168, 2004.

\bibitem[Rawlins and Culyer(2004)]{rawlins2004national}
Michael~D Rawlins and Anthony~J Culyer.
\newblock National {I}nstitute for {C}linical {E}xcellence and its value
  judgments.
\newblock \emph{BMJ: British Medical Journal}, 329\penalty0 (7459):\penalty0
  224, 2004.

\bibitem[Bostrom(2017)]{bostrom2017superintelligence}
Nick Bostrom.
\newblock \emph{Superintelligence}.
\newblock Dunod, 2017.

\bibitem[Sandberg and Bostrom(2008)]{sandberg2008whole}
Anders Sandberg and Nick Bostrom.
\newblock Whole brain emulation.
\newblock 2008.

\bibitem[Sandberg(2013)]{sandberg2013feasibility}
Anders Sandberg.
\newblock Feasibility of whole brain emulation.
\newblock In \emph{Philosophy and Theory of Artificial Intelligence}, pages
  251--264. Springer, 2013.

\bibitem[Jordan et~al.(2018)Jordan, Ippen, Helias, Kitayama, Sato, Igarashi,
  Diesmann, and Kunkel]{jordan2018extremely}
Jakob Jordan, Tammo Ippen, Moritz Helias, Itaru Kitayama, Mitsuhisa Sato, Jun
  Igarashi, Markus Diesmann, and Susanne Kunkel.
\newblock Extremely scalable spiking neuronal network simulation code: from
  laptops to exascale computers.
\newblock \emph{Frontiers in neuroinformatics}, 12:\penalty0 2, 2018.

\bibitem[Hanson(2016)]{hanson2016age}
Robin Hanson.
\newblock \emph{The Age of Em: Work, Love, and Life when Robots Rule the
  Earth}.
\newblock Oxford University Press, 2016.

\bibitem[Wootters and Zurek(1982)]{wootters1982single}
William~K Wootters and Wojciech~H Zurek.
\newblock A single quantum cannot be cloned.
\newblock \emph{Nature}, 299\penalty0 (~5886):\penalty0 802--803, 1982.

\bibitem[Dieks(1982)]{dieks1982communication}
DGBJ Dieks.
\newblock Communication by {EPR} devices.
\newblock \emph{Physics Letters A}, 92\penalty0 (6):\penalty0 271--272, 1982.

\bibitem[Price(2010)]{price2010decisions}
Huw Price.
\newblock Decisions, decisions, decisions: can {S}avage salvage {E}verettian
  probability?
\newblock In Simon Saunders, Jonathan Barrett, Adrian Kent, and David Wallace,
  editors, \emph{Many Worlds?: Everett, Quantum Theory, \& Reality}, pages
  369--391. OUP Oxford, 2010.

\bibitem[Kent(2010)]{kent2010one}
Adrian Kent.
\newblock One world versus many: the inadequacy of {E}verettian accounts of
  evolution, probability, and scientific confirmation.
\newblock In Simon Saunders, Jonathan Barrett, Adrian Kent, and David Wallace,
  editors, \emph{Many Worlds?: Everett, Quantum Theory, \& Reality}, pages
  307--354. OUP Oxford, 2010.

\bibitem[Reynolds and Ball(1963)]{reynolds1963little}
Malvina Reynolds and Ernie Ball.
\newblock \emph{Little {B}oxes}.
\newblock Essex Music of Australia Pty. Limited, 1963.

\bibitem[Tolstoy(1877)]{tolstoy1966anna}
Leo Tolstoy.
\newblock \emph{Anna {K}arenina}.
\newblock The Russian Messenger, 1877.

\end{thebibliography}

\end{document}